\documentclass[prl,twocolumn,superscriptaddress]{revtex4}

\usepackage{graphicx}
\usepackage{color}
\usepackage[usenames,dvipsnames,svgnames,table]{xcolor}
\usepackage{calc}
\usepackage{color}
\usepackage{amsmath,amssymb,amsthm}

\newif\ifprintcomments
\newif\iftest

\newcounter{comment}

\newif\ifsecheadings
\newcommand{\csection}[1]{\ifsecheadings\section{#1}\fi}
\newcommand{\csubsection}[1]{\ifsecheadings\subsection{#1}\fi}

\renewcommand{\vec}[1]{\mathbf{#1}}
\newcommand{\op}[1]{#1}

\newcommand{\I}{\mathrm{i}}

\newcommand{\bthet}{{\boldsymbol\theta}}

\printcommentsfalse

\testtrue

\secheadingsfalse

\begin{document}

\title{Two-dimensional Bloch electrons in perpendicular magnetic fields: 
an exact calculation of the Hofstadter butterfly spectrum}

\author{S.~Janecek$^\ast$} 
\affiliation{Instituto de Ciencia de Materiales de
  Madrid (ICMM--CSIC), Campus de Cantoblanco, 28047 Madrid, Spain}

\affiliation{Institut de Ciencia de Materials de
  Barcelona (ICMAB--CSIC), Campus de Bellaterra, 08193 Barcelona,
  Spain}

\affiliation{Johann Radon Institute for Computational and Applied
  Mathematics (RICAM), Austrian Academy of Sciences, Altenberger
  Strasse 69, A-4040 Linz, Austria} 

\affiliation{MathConsult GmbH, Altenberger
  Strasse 69, A-4040 Linz, Austria} 

\author{M. Aichinger}
  \affiliation{Johann Radon Institute for Computational and Applied
    Mathematics (RICAM), Austrian Academy of Sciences, Altenberger
    Strasse 69, A-4040 Linz, Austria} 
  \affiliation{Uni Software Plus GmbH, Kreuzstrasse 15a, A-4040 Linz, Austria}

\author{E.\;R.~Hern\'{a}ndez} 
\affiliation{Instituto de Ciencia de
  Materiales de Madrid (ICMM--CSIC), Campus de Cantoblanco, 28047
  Madrid, Spain} 
\affiliation{Institut de Ciencia de Materials de
  Barcelona (ICMAB--CSIC), Campus de Bellaterra, 08193 Barcelona,
  Spain}

\date{\today}

\begin{abstract} {The problem of two-dimensional, independent electrons
      subject to a periodic potential and a uniform perpendicular magnetic field unveils
      surprisingly rich physics, as epitomized by the fractal energy spectrum known as
      Hofstadter's Butterfly. It has hitherto been addressed using various approximations
      rooted in either the strong potential or the strong field limiting cases. Here we
      report calculations of the full spectrum of the single-particle Schr\"{o}dinger equation without
      further approximations.  Our method is exact,
      up to numerical precision, for any combination of potential and uniform field
      strength. We first study a situation that corresponds to the strong potential limit,
      and compare the exact results to the predictions of a Hofstadter-like model. We then
      go on to analyze the evolution of the fractal spectrum from a Landau-like
      nearly-free electron system to the Hofstadter tight-binding limit by tuning the
      amplitude of the modulation potential. }
\end{abstract}

\maketitle
%


The motion of electrons in a crystalline solid subject to a magnetic field has been
considered since the early days of quantum mechanics~\cite{Peierls:1933ZPh}. The
field splits the crystal's electronic bands  into sub-bands and internal minigaps; the
energetic arrangement of these sub-bands forms a fractal structure reminiscent of a
butterfly when plotted as a function of the field \cite{Hofstadter:1976wt}. Significant
experimental effort has been devoted to detecting signatures of this energy spectrum in
two-dimensional electron gases (2DEGs)~\cite{Albrecht:2001vs,Geisler:2004ei}.
For independent electrons, the situation is described by the single-particle
Schr\"{o}dinger equation,
\begin{equation}
  \label{eq:schroed}
  H \psi(\vec{r}) \equiv \left[ \frac{1}{2m} \Pi^2 + V(\vec{r})
  \right] \psi(\vec{r}) = E \psi(\vec{r}),
\end{equation}
where $H$ is the Hamiltonian, $\psi(\vec{r})$ is an eigenstate with
energy $E$, $\vec{\Pi} = \vec{p} + e \vec{A}(\vec{r})$ is the dynamical
momentum operator and $\vec{A}(\vec{r})$ is the vector potential
corresponding to the magnetic field, $\vec{B} = \nabla \times \vec{A}$. 
We take the field to be uniform and oriented along the $z$-direction,
$\vec{B} =B\, \vec{e}_z$. The electrons are restricted to the two-dimensional (2D)
$xy$-plane, and the external potential 
$V (\vec{r})$ is periodic on a Bravais lattice defined by vectors
\begin{equation}
  \label{eq:bravais}
  \vec{R}_\mathbf{n} = j \vec{a} + k \vec{b},\;\;
  \mathbf{n}= (j,k) \in \mathbb{Z}^2.
\end{equation}
To date, this problem has been chiefly approached by approximations starting from two
complementary limits, considering either the influence of a weak magnetic field on the
band structure resulting from a strongly varying potential $V(\vec{r})$ \cite{Harper:1955wk,
  Langbein:1969vg}, or the influence of a small modulation potential on the
Landau-quantized electrons in a strong field \cite{Wannier:1979wx, Langbein:1969vg}.  In
the strong potential limit, one typically starts with a tight-binding (TB) approximation
for a single band of the zero-field ($\vec{A}=0$) problem, $E(\vec{k})$, where $\vec{k}$
is the crystal momentum. Then, an effective Hamiltonian for the magnetic field problem is
generated from $E(\vec{k})$ through the Peierls substitution $\mathbf{k} \rightarrow
\left(\vec{p} + e \vec{A} \right)/\hbar$~\cite{Peierls:1933ZPh,Luttinger:1951ua}. This
procedure was used, among others, by Hofstadter in his seminal article for a nearest-neighbor (NN) TB
model of the 2D square lattice ~\cite{Hofstadter:1976wt}. There are a number of
simplifications inherent to this approach: (i) the TB approximation of the zero-field band
structure, (ii) the restriction to electrons in a single band, (iii) the neglect of the
diamagnetic energy of the TB orbitals and the field dependence of the TB hopping
integrals. This has been shown to lead to quantitative as well as qualitative errors for
both nearly free and tightly bound two-dimensional electrons~\cite{Hasegawa:1989wq,
  Nicopoulos:1990tw, Alexandrov:1991ts, Vugalter:2004ir}.  Generalization of the effective
Hamiltonian approach has turned out to be difficult, see, e.g., Ref.~\cite{Kohn:1959vm}.
Surprisingly, the strong field approach is closely related to the strong potential one: if
potential-induced coupling between different Landau levels is neglected, the same secular
equation results, but with the magnetic field replaced by its inverse
\cite{Langbein:1969vg}.  Including such coupling has a profound effect on the calculated
energy spectrum \cite{Springsguth:1997vo, Kuhn:1993tq}, even for weak coupling
strength. The resulting rearrangement of the Hofstadter butterfly has recently been confirmed
by experiments \cite{Geisler:2004ei}; we take this as a strong indication that experiments
can only be fully understood by going beyond the approximate schemes described above.

\csection{Method}

\paragraph{Method.} In the absence of magnetic field, Bloch's theorem, which results from
the commutation of the lattice translations with the Hamiltonian, allows to restrict the
calculation of the eigenfunctions $\psi(\vec{r})$ to one primitive cell of the lattice in
Eq.~\eqref{eq:bravais}. As these translations do not leave the vector potential
$\vec{A}(\vec{r})$ invariant, they no longer commute with $H$ when a magnetic field is
present; consequently, the eigenfunctions $\psi(\vec{r})$ \emph{are not Bloch waves}.  By
combining a lattice translation with a suitable gauge transformation to counteract its
effect upon the vector potential, one can define {\em magnetic translation operators\/},
$\op{T}_A(\vec{R}_\vec{n})$, that do commute with $H$, see Refs.~\cite{Brown:1964vc,
  Zak:1964vu, Fischbeck:1970vz}. The operators $\op{T}_A(\vec{R}_\vec{n})$ do not, however,
form a group. For a rational field, where the number of magnetic flux quanta per unit
cell, $\alpha = \frac{1}{2\pi} \frac{e}{\hbar} \left(\vec{a} \times \vec{b} \right)B$, is
a rational number, $\alpha=p/q$ with $p$ and $q$ relative prime, one can choose a larger
magnetic lattice that has an integer number of $p$ flux quanta passing through each cell,
e.g.,
\begin{equation}
  \label{eq:muc1}
  \vec{S}_\mathbf{n} = j \vec{a} + k (q \vec{b}),\;\;
  \mathbf{n}= (j,k) \in \mathbb{Z}^2.
\end{equation}
On this lattice, a generalized version of the Bloch theorem holds~\cite{Fischbeck:1970vz}:
for an orthorhombic lattice and Landau gauge, $\vec{A}(\vec{r})=Bx\vec{e}_y$, the
solutions of Eq.~\eqref{eq:schroed} can be chosen of the form $\phi(\vec{r}) \equiv e^{\I
  \bthet\vec{r}} u^\bthet(\vec{r})$, where $u^\bthet(\vec{r})$ obeys
\begin{equation}
  \label{eq:genbloch1}
  u^\bthet(\vec{r}+\vec{S}) = \exp\left[ - \I \frac{e}{\hbar}
   B S_x y \right]  u^\bthet(\vec{r}),
\end{equation}
and the magnetic crystal momentum $\bthet$ is restricted to the first Brillouin zone of
\eqref{eq:muc1}. This condition allows to restrict the calculation of $u^\bthet(\vec{r})$
to one primitive cell of the magnetic lattice. The functions $u^\bthet(\vec{r})$ exhibit a
peculiar topology in a space where their $y$-component is expanded in a Fourier series,
see Ref.~\cite{Cai:2004ht}. This allows to take into account the boundary condition
\eqref{eq:genbloch1} in a natural way when the diffusion method is used for solving the
Schr\"odinger equation~\cite{Chin:2009ig, Aichinger:2005ik}. We have summarized the steps
leading to Eq.~\eqref{eq:genbloch1} and the technical details of the diffusion method in
the Supplementary Information.

At a fixed rational field, $\alpha=p/q$, we obtain the bands $E_j(\bthet)$ by numerically
solving Eq.~\eqref{eq:schroed} for a grid of $\bthet$-values spanning the magnetic
Brillouin zone (MBZ) corresponding to this field.  The density of states (DOS) as a
function of the field, $\rho(B,E)$, can then be calculated by integrating the bands over
the MBZ and repeating the process for different fields. The rational field can only be
tuned in discrete steps, $\alpha_P = P/Q$, where $P,Q \in \mathbb{Z}$.  The size of the
magnetic unit cell depends on the reduction of $P/Q$ to a quotient of relatively prime
integers $p/q$, and is $q$ times as large as the zero-field (``geometric'') unit cell. At
the field $\alpha_P$, one zero-field band is thus expected to split into $q$ magnetic
bands, which are known to cluster in groups of $p$ bands~\cite{Hofstadter:1976wt}.  The
pattern of distinct prime factors of $P$ and $Q$ as $\alpha_P$ is swept across a range of
fields gives rise to the self-similar, fractal pattern of gaps in the DOS that has become
known as the Hofstadter butterfly.

When the Fermi energy of the system lies in a gap, i.e., a region where $\rho(B,E_F)$ is
zero, the Hall conductance $\sigma_{xy}$ assumes a quantized value,
$\sigma_{xy}^\mathrm{gap} = n e^2/h, n \in \mathbb{Z}$.  The fundamental topological
reason for this quantization was revealed by Thouless \emph{et
  al.}~\cite{Thouless:1982wi}, who showed that both in the strong field and strong
potential limits the Kubo-Greenwood formula for $\sigma_{xy}$ is related to the Chern
number of the $U(1)$ bundle over the magnetic Brillouin zone. This was later argued
\cite{Dana:1985vg} to be a direct consequence of the magnetic translation symmetry,
Eq.~\eqref{eq:genbloch1}.  Sweeping either the magnetic field or the Fermi energy through the
fractal pattern of minigaps inside a broadened band or Landau level results in a peculiar,
non-monotonous Hall effect~\cite{Thouless:1982wi}. Indications of this behavior have been
found in experiments \cite{Albrecht:2001vs}.  In this work we have used an alternative
approach, introduced by St\v{r}eda~\cite{Streda:1982ve}, to obtain
$\sigma_{xy}^\mathrm{gap}$ from the numerically calculated DOS,
\begin{equation}
  \label{eq:streda}
  \sigma_{xy}^\mathrm{gap}(B,E_F) = e \frac{\partial \rho(B,E')}{\partial B}
  \Big|_{E'=E_F}.
\end{equation}

\csection{Results and Discussion}

\begin{figure}[tbh]
  \centering
 \includegraphics[width=3.2in, keepaspectratio]{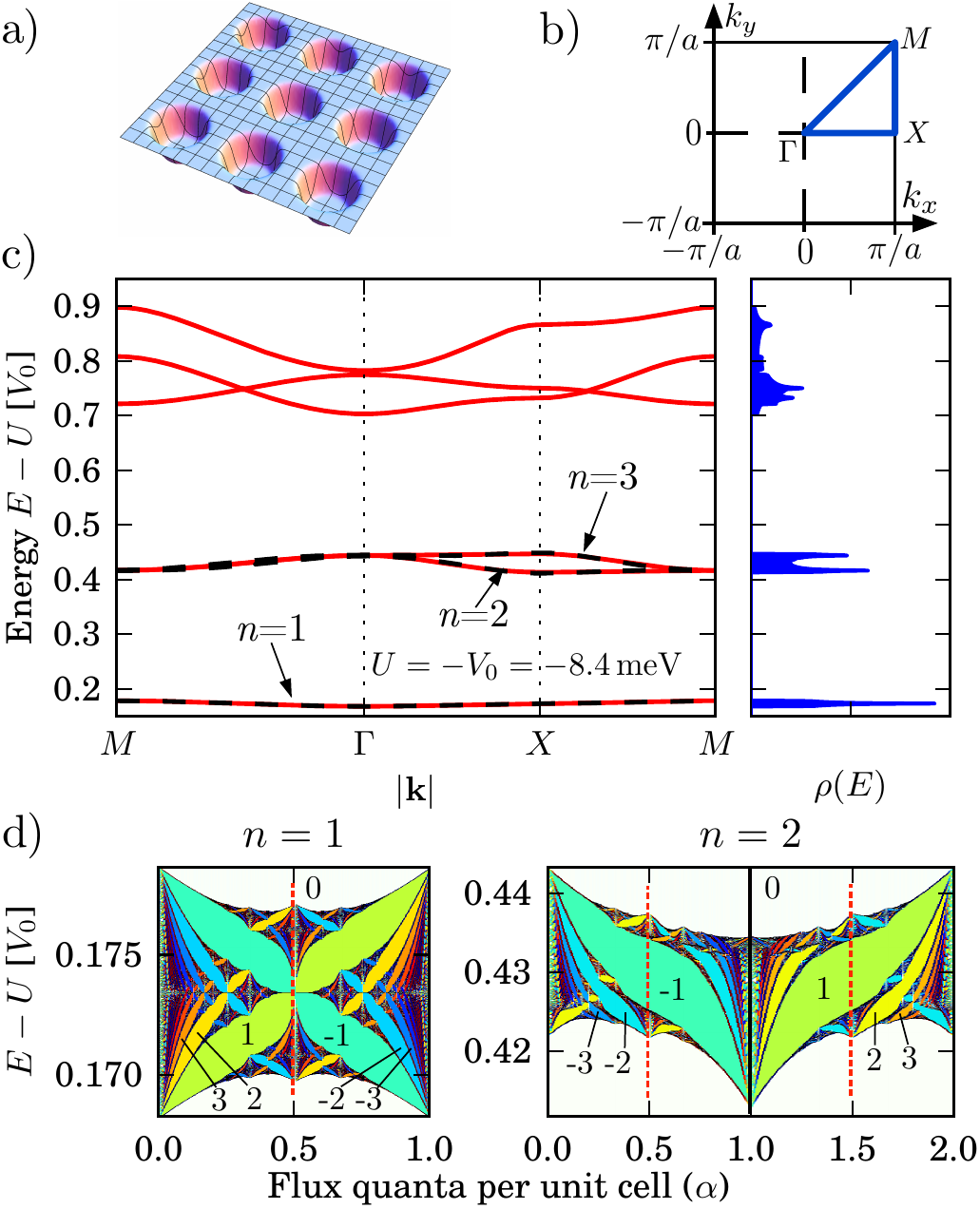}
 \caption{(color online). a) Schematic plot of the periodic Fermi well potential.  b) First Brillouin zone
   of the reciprocal lattice. c) \emph{Left panel}: Zero-field band structure of the
   potential with parameters \eqref{eq:params} and $U=-V_0$. Solid red lines show bands
   obtained by numerically solving the Schr\"odinger equation, dashed black lines are TB
   bands fitted to these exact bands (see text). \emph{Right panel}: DOS $\rho(E)$
   obtained from the full Schr\"odinger equation.  d) Hofstadter butterflies for the two lowest
   bands.  Areas with non-zero DOS are printed black, and the gaps are colored according
   to the corresponding quantized Hall conductance $\sigma_{xy}^\mathrm{gap}$ in units of
   $e^2/h$. White indicates zero Hall conductance, warm (cold) colors indicate positive
   (negative) Hall conductance. A number of larger gaps are labeled with the
     corresponding Hall conductance for reference. The butterflies are periodic in the
   flux, one period being shown in each case.  }
  \label{fig:bands_dos}
\end{figure}

In our numerical calculations, we have studied a simple system consisting of a
two-dimensional square lattice of potential wells with the symmetrized Fermi function
form~\cite{Sprung:1997vk},
\begin{equation}
  \label{eq:symmfermi}
  V(r) = U  \coth \left(\frac{r_0}{2 d}\right) \frac{ \sinh \left(\frac{r_0}{d}\right)}{\cosh
    \left(\frac{r}{d}\right)+\cosh \left(\frac{r_0}{d}\right)},
\end{equation}
with parameters
\begin{equation}
  \label{eq:params}
  r_0 = 39.7\,\text{nm},\; d = 1.59\,\text{nm},\; a = 100\,\text{nm},
\end{equation}
where $a$ is the lattice spacing. The potential is illustrated in
Fig.~(\ref{fig:bands_dos}a); the parameters are set to loosely reproduce the conditions of
earlier experimental studies~\cite{Albrecht:2001vs,Geisler:2004ei}. 

\csubsection{Strong Potential Case}

\paragraph{Strong Potential Regime.} 
We first compare the spectrum generated by the full Schr\"odinger
equation~\eqref{eq:schroed} to the results of a TB approximation similar to the
one used by Hofstadter.  We thus choose a fairly deep modulation potential,
\begin{equation}
  \label{eq:utb}
  U=-V_0 \equiv -8.4\,\text{meV}.
\end{equation}
We obtained the band structure and DOS at zero magnetic field [see
Fig.~(\ref{fig:bands_dos}c)] by numerically solving the corresponding single-particle
Schr\"odinger equation.  
We then fitted a TB model to the lowest three bands [black dashed
lines in Fig.~(\ref{fig:bands_dos}c)], and employed the Peierls substitution to obtain the
fractal energy spectra shown in Fig.~(\ref{fig:bands_dos}d), see the Supplementary Information for
details. The lowest band [$n=1$ in Fig.~(\ref{fig:bands_dos}c)] required only
nearest-neighbor hopping integrals in the TB band model for an adequate fit, and thus
yields a spectrum corresponding to that obtained by Hofstadter~\cite{Hofstadter:1976wt}
[left panel of Fig.~(\ref{fig:bands_dos}d)]. The second and third bands required up to
$3^{rd}$-nearest-neighbor hoppings, which lead to significantly distorted versions of
Hofstadter's butterfly, shown for the second band on the right panel of
Fig.~(\ref{fig:bands_dos}d). The spectrum of the third band is similar and is not
shown. Our findings qualitatively agree with the results of Ref.~\cite{Claro:1981vl}.

\begin{figure}[htbp]
  \centering
  \includegraphics[width=3.4in, keepaspectratio]{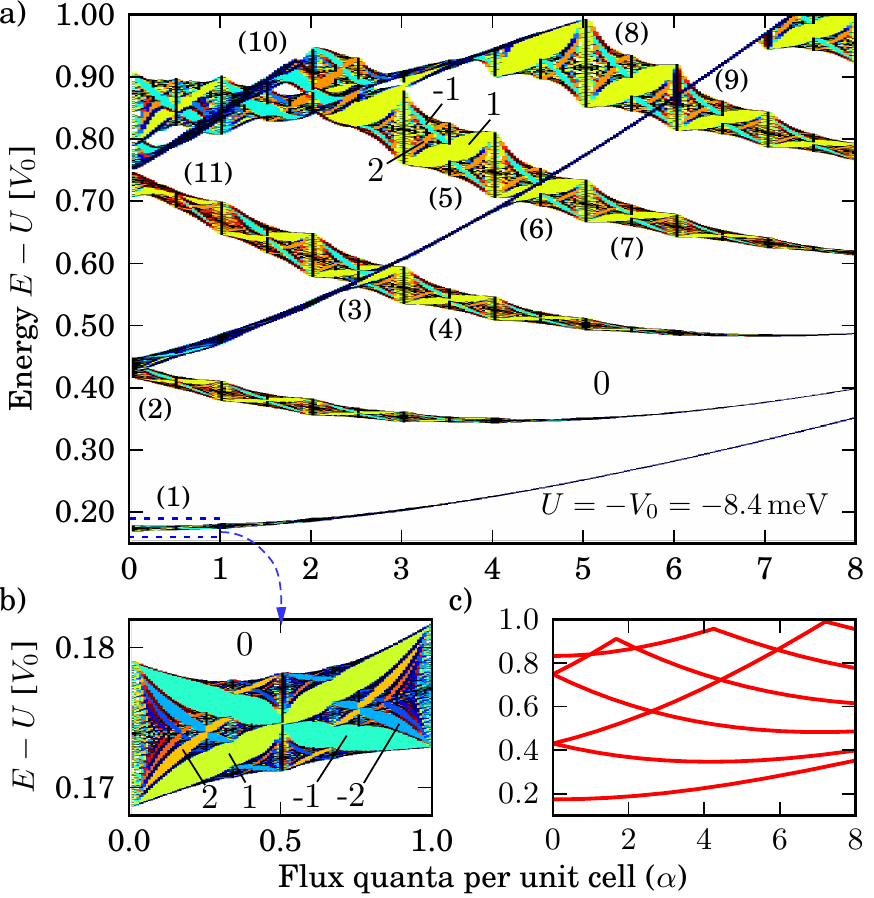}
  \caption{(color online). Magnetic energy spectrum (DOS and Hall conductance $\sigma_{xy}^\mathrm{gap}$)
    for the lowest six bands of the of the square Fermi well lattice with parameters
    \eqref{eq:params} and $U=-V_0$, calculated by numerically solving
    Eq.~\eqref{eq:schroed}. The color coding and labeling is as in
    Fig.~(\ref{fig:bands_dos}).  Panel ({\em b\/}) shows a magnified portion of the lowest
    band indicated by the dashed box in panel ({\em a\/}).  Panel ({\em c\/}) shows the
    energy spectrum of an isolated Fermi well potential, Eq.~(\ref{eq:symmfermi}). It
    strongly resembles the spectrum of a parabolic potential (the Fock-Darwin
    spectrum~\cite{FockDarwinDarwin}).}
  \label{fig:sbat_six}
\end{figure}

In Fig.~(\ref{fig:sbat_six}a) we show the energy spectrum of the six lowest bands, as
obtained by numerically solving the full magnetic eigenvalue problem,
Eq.~\eqref{eq:schroed}, using the scheme outlined above.
A maximum magnetic unit cell size of $Q=32$ was employed in the calculation. For
comparison, the spectrum of an isolated potential well is plotted in
Fig.~(\ref{fig:sbat_six}c). It bears a strong resemblance to the Fock-Darwin~(FD)
spectrum~\cite{FockDarwinDarwin} of a parabolic well, we will thus refer to these states
as ``FD states'' in the following.  It can be seen in Fig.~(\ref{fig:sbat_six}a) that in
the periodic system the FD states of the isolated well are broadened into bands with a
fractal internal structure that is qualitatively well described by the Hofstadter
butterfly. In the exact result the periodicity of the Hofstadter spectrum is superimposed
onto the field dependence of the corresponding FD state; this field-dependence is not
taken into account in the TB model with constant hopping integrals.  In general, higher
energy levels, having more extended wave functions, undergo larger broadening at a given
flux value. Conversely, bands become narrower with increasing field, as their wave
functions become more spatially localized, tending to Landau levels in the limit of high
field intensities. In the TB model, the main effect of increasing the second- and
third-nearest neighbor hoppings is a distortion of the butterfly that opens a gap at flux
strengths of $\alpha_j =j+1/2$, with $j \in \mathbb{Z}$, indicated by the vertical (red)
dotted lines in the butterflies in Fig.~(\ref{fig:bands_dos}d). This behavior is also
present in the exact results, compare regions (4)--(5) or (7)--(8) in
Fig.~(\ref{fig:sbat_six}a).  The gap widens for higher-energy bands, consistent with the
TB approximation, where such bands need to be modeled by larger hopping integrals to more
distant neighbors.  The gap decreases again with increasing field due to the stronger
localization of the wave functions, an effect that is not included in the TB description
with $B$-independent hopping. Band crossings, which are not described within the
single-band Peierls approximation, are an interesting subject for future studies: the
narrow third FD band seems to disrupt the butterfly pattern of the broader bands it
crosses in regions (3) and (6), but does not seem to exert any noticeable effect in region
(9). At higher energies [see region (10)], where multiple bands cross, the resulting
fractal spectrum can assume a form that is very different from the original Hofstadter
butterfly.

\csubsection{Intermediate Potential Regime}

\begin{figure*}[hbtp]
  \centering
  \includegraphics[width=7in,keepaspectratio]{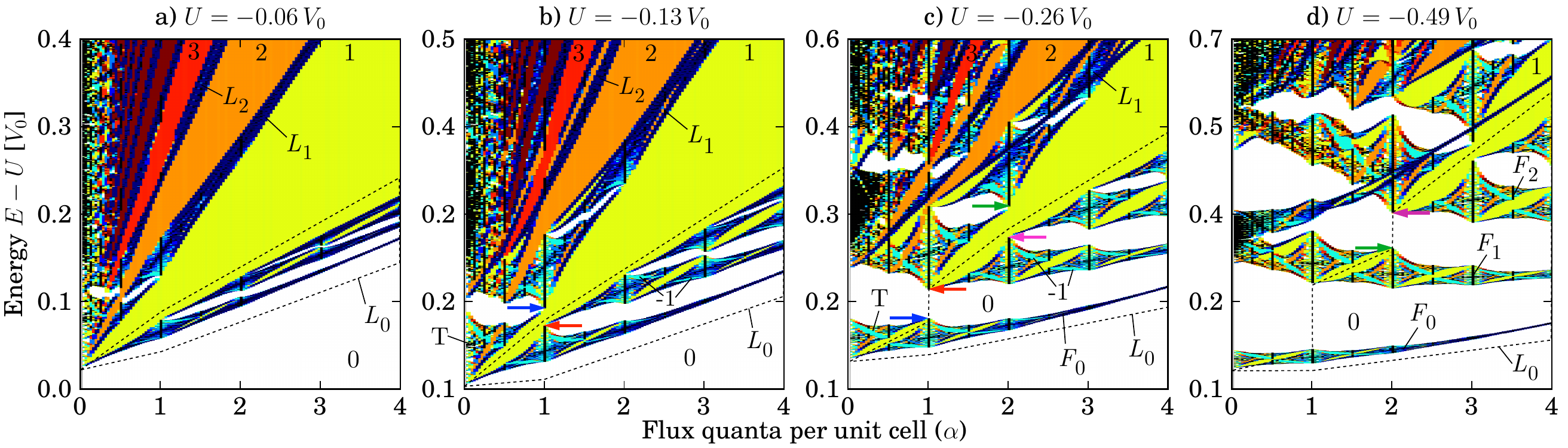}
  \caption{(color online). Energy spectra for different strengths of the modulation potential, ranging
    from a ``Landau-like'' system of nearly free electrons [$U=-0.06\, V_0$, panel a)] to
    a moderately deep potential [$U=-0.49\,V_0$, panel d)] already resulting in a
    ``Fock-Darwin''-like spectrum. Energies and modulation potentials $U$ are given in
    units of $V_0$, the depth of the modulation potential used to generate
    Fig.~\ref{fig:sbat_six}.  Gaps are color coded an labeled with their corresponding
    Hall conductance as in Fig.~\ref{fig:bands_dos}. Landau levels are denoted by
    $L_1,L_2,\dots$, Fock-Darwin-like states $F_0, F_1, \dots$. The dotted line encloses
    the broadened lowest Landau level. The label ``T'' indicates the triangular cluster of
    states referenced in the text.}
  \label{fig:sbatmovie}
\end{figure*}

\paragraph{Intermediate Potential Regime.} We now explore the evolution of
the spectrum as the potential is changed from nearly flat (strong field limit) to highly
modulated (strong potential limit). The intermediate stages of this evolution are not
accessible to the approximate methodologies hitherto employed. We use the square lattice
potential with the parameters given in Eq.~\eqref{eq:params}, but change the well
depth~$U$.  We start with a very shallow potential ($U =- 0.06\, V_0$), which corresponds
to a Landau-like system with nearly-free electrons in a magnetic field. The resulting
spectrum is illustrated in Fig.~(\ref{fig:sbatmovie}a); it exhibits the typical ``Landau
fan'' form, with slightly broadened Landau levels (LLs) that display an internal fractal
structure of minigaps (most evident in the lowest level, $L_0$).
The Hall conductance between LLs increases monotonically in steps of
$e^2/h$, consistently with the integer quantum Hall effect. Furthermore, an emerging white
gap ($\sigma_{xy}^\mathrm{gap}=0$) can be observed at low field ($\alpha=0-1$, $E \approx
0.11\,V_0$).  When the potential modulation is increased to $U = -0.13\,V_0$
[Fig.~(\ref{fig:sbatmovie}b)] this gap is seen to widen further, practically cutting off a
low-energy triangular section (marked ``T'' in the figure) from all LLs $L_n$ with
$n>0$. At $\alpha = 1$ the upper-right tip of this triangular cluster of states retains a
tenuous link to the broadened $L_1$ Landau band, at the position indicated by the blue
arrow. At the same time, the minigaps at $\alpha \geq 1$ in the lowest Landau band $L_0$
have broadened further, to the point where the band is only held together by a narrow
sub-band at $\alpha = 1$, indicated by the red arrow.  The Hall conductance in the gap
between $L_0$ and $L_1$ at $\alpha=1$ is still $\sigma_{xy}^\mathrm{gap}=1$. But upon
further increasing the modulation strength, the gap first closes and then reopens with the
above links reversed: at $U = -0.26\,V_0$ [Fig.~(\ref{fig:sbatmovie}c)], the tip of the
triangular cluster is now connected to the lowest miniband of $L_0$ (blue arrow), while
the rest of the $L_0$ miniband is connected to $L_1$ (red arrow). At the same time, the
gap has changed its character to $\sigma_{xy}^\mathrm{gap}=0$, resulting in the first
clearly discernible FD-like band $F_0$, separated from all higher energy states by an
unbroken white gap. Most of $F_0$ is formed from the lowest miniband of the LL $L_0$,
except for the triangular cluster of states in $\alpha<1$, which originates from higher
LLs.  The process of formation of the first FD band ($F_0$) that is incipient at $\alpha =
1$ in Fig.~(\ref{fig:sbatmovie}b) can be seen to repeat itself at $\alpha = 2$ for the
second FD band ($F_1$) [Fig.~(\ref{fig:sbatmovie}c), green and purple arrows] and at
$\alpha = 3$ for the third FD band ($F_2$) [Fig.~(\ref{fig:sbatmovie}d)]. The gradual
transformation from LLs to FD-like bands thus proceeds by the re-connection of minibands
from one broadened LL to those of neighboring LLs. This rearrangement process fragments
the large triangular gaps between LLs in the weak potential limit into minigaps
encapsulated by the emerging FD-like bands. As a result, the FD states $F_0$, $F_1$, $F_2$
in Fig.~\ref{fig:sbatmovie}d are composed of a low-field section that originates from
higher LLs (triangle T in case of $F_0$, one Hofstadter butterfly segment plus a
triangular cluster in case of $F_1$, etc.), and a high-field section that is one miniband
of the lowest LL $L_0$.

\csection{Conclusion}

The above results show that with the method presented here it is possible to perform exact
calculations of the spectrum of independent electrons in a 2D~periodic potential and
constant perpendicular magnetic field.  The same technique is readily applicable to solve
the Kohn-Sham equations of density functional theory~(DFT) which are expected to provide a
reasonable description at least for weakly correlated electron systems.  Experimental
techniques have recently become available to directly probe the local DOS in 2DEGs on
surfaces in a perpendicular magnetic field using scanning tunneling spectroscopy, making
possible the detection of spatial features of some LLs~\cite{Hashimoto:2012}, and even
measuring their response to a 1D~periodic potential due to surface
buckling~\cite{Okada:2012}. Such developments may presently bring about additional
possibilities of experimentally determining the spectral features of the 2DEG subject
simultaneously to a periodic potential and a perpendicular magnetic field, contrasting
them with the predictions reported herein. The Hofstadter butterfly is not unique to the
system discussed here. The basic principle underlying the fractal energy spectrum is the
presence of two competing symmetries (in the case studied here, the periodicity of the
lattice and the symmetry of the Landau orbits, which is governed by the area required by
one magnetic flux quantum).  Similar patterns have been observed or predicted to occur in a
variety of very different systems, such as microwaves transmitted through a waveguide with
a periodic arrangement of scatterers~\cite{Kuhl:1998wp}, the
electronic~\cite{deLange:1983} and vibrational~\cite{Quilichini:1997wu} spectra of
incommensurate crystals, ultracold atoms in optical lattices~\cite{Jaksch:2003}, and
photonic crystals \cite{Fang:2012js, Hafezi:2011dt}.  The topological protection of the
quantum Hall phase has been shown to improve the performance of optical delay lines and to
overcome limitations related to disorder in photonic technologies \cite{Hafezi:2011dt}. We
hope that the method presented here will also prove beneficial in these related fields.

We wish to thank A.\ Garc\'{\i}a and E.\ Krotscheck for helpful discussions.  SJ was
funded by the Austrian Science fund FWF under project no.~J2936-N, E.R.H. by the Spanish
Ministry of Science and Innovation through project FIS2009-12721-C04-03. SJ would like to
thank ICMAB and ICMM for their hospitality during his stay. We acknowledge CESGA and the
Johannes Kepler University Linz for the use of their computer facilities, where the
results reported here were obtained.

%
\bibliographystyle{apsrev}
\end{document}